\documentclass[a4paper,12pt]{article}

\newcommand{\J}{\mathrm{i}}  
\newcommand{\D}{\mathrm{d}}  
\newcommand{\E}{\mathrm{e}}  

\newcommand{\dodoi}[1]{doi:#1}

\usepackage{graphicx}
\usepackage{amsmath}
\usepackage[squaren]{SIunits}

\graphicspath{{./Figures/}}

\begin{document}

\title{Sound from rotors in non-uniform flow}

\author{Michael Carley}






\maketitle

\begin{abstract}
  An analysis is presented for the evaluation of the acoustic field of
  a rotating source in a non-uniform potential flow. Other than the
  restriction to low flow Mach numbers, the method is exact and
  general. The variation in radiation properties with source position
  is handled by representation as a Fourier series in source angle,
  giving rise to an asymmetrically varying acoustic field evaluated by
  summation of the series. The method is used to develop an exact
  solution for the model problem of a rotor operating near a cylinder
  in cross-flow and sample calculations demonstrate the accuracy of
  the technique when compared to full numerical evaluation. The
  calculations show changes of greater than one decibel in the
  acoustic field due to flow non-uniformity at a flow Mach number
  of~0.15, a typical speed for propeller aircraft at take-off.
\end{abstract}


\maketitle


\section{Introduction}
\label{sec:intro}

The prediction and analysis of sound from rotating sources is a
long-standing problem in acoustics, for its theoretical interest and
for its importance in many industrial applications. The problem can be
viewed as made up of the determination of the source terms, typically
related to the fluid dynamics of the system, and to the propagation of
the resulting sound. In this paper, we consider propagation of sound
from a known source, on the assumption that the source can be
determined by some other means.

Even on linear theory, the calculation of the sound radiated by a
rotating source, here modeled as a sinusoidally-varying source
distributed on a ring, contains many features which complicate the
calculation. The development of prediction methods dates back some
hundred years to the early studies of noise from aircraft
propellers~\cite{lynam-webb19}. In this work, the source is assumed to
be at rest and to radiate into a stationary fluid. The next extension
to the theory was the inclusion of flow effects in the form of a
uniform flow parallel to the axis of the
source~\cite{gutin48,garrick-watkins53}, corresponding to a propeller
in forward flight, or to a propeller fixed in a mean flow, as in a
wind tunnel. The introduction of flow generates effects not present in
the stationary fluid case, because of changes in radiation efficiency
caused by the increased speed of the source relative to the fluid, and
convective effects on retarded time. The acoustic field, however,
remains symmetric about the axis of rotation of the source, because
of the axial symmetry of the flow. 

The next advance in the theoretical treatment of the problem was the
introduction of flow which is not symmetric about the source axis. For
a uniform flow this arises when a propeller is inclined at some angle
to the incoming flow, for example during
take-off~\cite{stuff88,mani90,hanson95,campos99}. In this case,
although the flow is uniform, because it is not aligned with the axis
of rotation of the source, the radiation efficiency varies as a
function of source angle. Early studies concentrated on the
flow-generated unsteady source terms, which were known to be efficient
radiators of sound and only later were asymmetric convection effects
considered. Even if we consider only a steady source, of constant
strength in its frame of reference, this variation in radiation
efficiency gives rise to an asymmetric field. The asymmetry can be
modeled in a number of ways. In this paper, we adopt the viewpoint of
Hanson~\cite{hanson95} who predicts the acoustic field as a Fourier
series in azimuthal angle about the source axis, with the source also
decomposed into a Fourier series or azimuthal modes. When there is no
flow, or the flow is not inclined with respect to the rotor axis, each
source mode gives rise to one field mode of the same order. When the
flow is inclined with respect to the source axis, a single source mode
generates multiple field modes, whose magnitudes depend on the source
characteristics, and on the angle and speed of the flow.

In this paper, we extend the analysis of radiation from ring sources
to the case where the source radiates into a non-uniform flow, a
generalization from the existing theories for uniform flow. The only
limitation is that the theory used is correct to first order in flow
Mach number. In underwater applications this is not a significant
restriction, but it does mean that the full range of speeds relevant
to aeronautical applications is not covered. In practice, however,
noise from aircraft is most troublesome during take-off and landing
when aircraft are near the ground, and in these cases aircraft are
typically operating at low flight speed, so the restriction to low
Mach number is not as limiting in practice as it might appear.

The rest of the paper begins by introducing existing analytical tools,
for the prediction of sound from rotating sources, and for the
inclusion of low-speed non-uniform flow effects, respectively, before
combining them into the main result of the paper, a method for the
prediction of the acoustic field from rotating sources in non-uniform
flow. A model problem is then introduced, for sound from a rotor near
a cylinder in cross-flow, and sample results are presented to
demonstrate the performance of the method and the size of acoustic
effect induced by non-uniform flow.

\section{Analysis}
\label{sec:analysis}

Analysis of the problem of radiation from rotating, or ring, sources
is handled by treating the source as having sinusoidal variation in
azimuthal angle around the ring. Each such term gives rise to an
integral which can be expressed exactly by a series expansion and, if
necessary, the contribution from terms can be summed to give the
overall field at any given frequency. In this paper, we follow the
same procedure, but we must take account of the azimuthal variation in
radiation properties caused by the non-uniform flow. This is done by
representing the variation in flow potential as a set of azimuthal
modes which convert the problem of radiation in non-uniform flow into
a sum of modes radiating into a static fluid.

In the rest of the paper, we do not consider the determination of the
magnitude of source terms, typically found from the aerodynamics of
the problem, nor do we extend the analysis to include interaction of
the radiated field with bodies in the flow. This is consistent with
previous analyses~\cite{stuff88,mani90,hanson95} which have looked
only at the effects of flow on direct radiation from the source,
without considering the secondary effect of scattering by nearby
bodies.

\subsection{Sound from ring sources}
\label{sec:analysis:ring}

\begin{figure}
  \centering
  \includegraphics{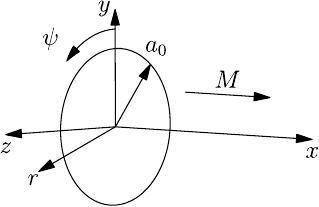}
  \caption{Geometry and notation for radiation from ring source of
    radius $a_{0}$ in cylindrical coordinate system $(r,\psi,x)$
    centered on ring}
  \label{fig:analysis:ring}
\end{figure}

The first step in evaluating the field from a ring source is a
standard integral which has been studied extensively for many
years~\cite{gutin48,wright69,stuff88,hanson95,carley10}.
Figure~\ref{fig:analysis:ring} shows the geometry and notation. We
adopt cylindrical coordinates $(r,\psi,x)$ with the source lying on a
ring of radius $a_{0}$ in the plane $x=0$. The source has strength
$s(\psi)\exp[-\J\omega t]$, with $\omega$ frequency and $t$ time.  The
time variation is suppressed from this point onwards. The acoustic
potential at a field point $(r,\psi,x)$ is given by integration on the
source ring,
\begin{align}
  \phi(r,\psi,x) 
  &=
  \int_{0}^{2\pi}
  \frac{\E^{\J k R}}{4\pi R}
  s(\psi_{1})
  \,
  \D\psi_{1},\\
  R^{2} &= r^{2} + x^{2} + a_{0}^{2} - 2a_{0}r\cos(\psi - \psi_{1}),
\end{align}
with $c$ speed of sound and wavenumber $k=\omega/c$. Subscript~$1$
denotes a variable of integration. Decomposing the source term into a
Fourier series in azimuthal angle $\psi_{1}$, 
\begin{align}
  s(\psi_{1}) &= \sum_{n=-\infty}^{\infty} s_{n}\E^{\J n\psi_{1}},
\end{align}
the potential is rewritten as a sum of terms
\begin{align}
  \phi(r,\psi,x)
  &=
  \sum_{n=-\infty}^{\infty}
  s_{n}
  \E^{\J n\psi}
  I_{n}(k, M, a_{0}; r, x).
\end{align}
The integral $I_{n}(k, M, a_{0}; r, x)$ is the acoustic potential
radiated by a single azimuthal mode of the source in a flow of Mach
number $M$, and for $M=0$,
\begin{align}
  \label{equ:analysis:ring}
  I_{n}(k,0,a_{0}; r, x)
  &=
  \int_{0}^{2\pi}
  \frac{\E^{\J (k R_{1} + n\psi_{1})}}{4\pi R_{1}}
  \,\D\psi_{1},\\
  R_{1}^{2} &= x^{2} + r^{2} -2a_{0}r\cos\psi_{1} + a_{0}^2,
\end{align}
and is taken as the quantity of interest in the rest of the paper. The
integral has an exact series representation which will be used later
to calculate the acoustic field~\cite{carley10},
\begin{align}
  \label{equ:analysis:series}
  I_{n}(k,0,a_{0}; r, x)
  &=
  \frac{\J^{2n+1}}{2}
  \sum_{q=0}^{\infty}
  v_{q}
  h_{n+2q}^{(1)}(k\rho)j_{n+2q}(k
  a_{0})P_{n+2q}^{n}(x/\rho),\\
  v_{q} &=   (-1)^{q}
  (2n+4q+1)
  \frac{(2q-1)!!}{(2n+2q)!!},\\
  \rho^{2} &= x^{2} + r^{2},\quad \rho>a_{0}, n\geq 0,
\end{align}
where $(\cdot)!!$ is the double factorial, $h_{n}^{(1)}(\cdot)$ is the
spherical Hankel function of the first kind, $j_{n}(\cdot)$ is the
spherical Bessel function, and $P_{n}^{m}(\cdot)$ is the associated
Legendre function. For negative $n$, the field can be evaluated using
$I_{n}=I_{-n}$. 

The form of the acoustic field generated by the ring source is
controlled by the parameter $M_{t}=ka_{0}/n$. For a steady source
rotating in a uniform flow, this is the Mach number associated with
the rotation of the source, and arises as a parameter in the analysis
of propeller noise where it corresponds to the velocity at the ``tip''
of a rotor blade. When asymmetry is introduced into the system, this
gives rise to modes of varying order $n$, at constant wavenumber
$k$. The parameter also arises in Tyler and Sofrin's classic study of
propagation in ducts~\cite{tyler-sofrin62} where the effect of
changing the mode order is described in terms of changes to
$M_{t}$. In particular, the propagation properties change when
$M_{t}>1$, the case of a ``supersonic mode''. This happens when a
non-uniformity of some kind generates extra modes of order $n\pm m$,
$m=1,2,3,\ldots$. Should there be a value of $m$ such that $m\geq
n-ka_{0}$, $M_{t}\geq 1$ and the $n-m$ mode can radiate
efficiently. If the amplitude of this mode is large enough, this can
lead to large changes in the acoustic field.

The remainder of this paper extends the analysis for a ring source to
the problem of radiation into non-uniform flow by converting it into a
sum of integrals of the form of Equation~\ref{equ:analysis:ring},
using Taylor's transformation for propagation in low Mach number
potential flow~\cite{taylor78}.

\subsection{Sound propagation in non-uniform flows}
\label{sec:analysis:taylor}

Taylor's transformation~\cite{taylor78,taylor79,agarwal-dowling07} is
a method for converting problems of propagation in non-uniform flows
at low Mach number into equivalent problems in a stationary fluid, by
modifying the phase term of the Green's function as a function of the
position $\mathbf{x}$ and flow velocity. The transformation was
originally derived\cite{taylor78} using a coordinate transformation of
the wave equation with non-uniform flow, but can also be derived from
the geometrical acoustics approximation, as shown in the Appendix.

The flow velocity is given by a velocity potential
$cM\Phi(\mathbf{x})$ where $M$ is the flow Mach number and
$\Phi(\mathbf{x})$ is the velocity potential for unit velocity and
multiplication by $cM$ scales the flow field to the required velocity.
To first order in $M$, the acoustic potential at a point $\mathbf{x}$
due to a point source at $\mathbf{x}_{0}$ is given by
\begin{align}
  \label{equ:analysis:taylor}
  \phi(\mathbf{x}) &= \frac{\E^{\J k R}}{4\pi R}
  \E^{\J kM\Phi(\mathbf{x}_{0})}
  \E^{-\J kM\Phi(\mathbf{x})}.
\end{align}
The prediction of sound in a low Mach number non-uniform flow thus
becomes a phase-shifted version of the problem in a stationary
fluid. The difficulty of the ring source problem lies in representing
this phase shift in the radiation integral of
Equation~\ref{equ:analysis:ring} in a manner amenable to analysis.

\subsection{Sound from ring sources in non-uniform flows}
\label{sec:analysis:non:uniform}

Inserting Taylor's transformation, Equation~\ref{equ:analysis:taylor},
into Equation~\ref{equ:analysis:ring} yields an integral
representation for the acoustic potential in a general potential flow
at low Mach number. The integral can be evaluated numerically but this
may be time-consuming and yields no physical insight. We now show how
the phase term of the Taylor transformation can be represented as a
set of azimuthal modes converting the integral of
Equation~\ref{equ:analysis:ring} into a sum of modes radiating into a
stationary fluid.

Upon substitution of the Taylor transformation into
Equation~\ref{equ:analysis:ring}, the radiation integral becomes
\begin{align}
  \label{equ:analysis:taylor:int}
  I_{n}(k, M, a_{0}; r, x)
  &=
  \E^{-\J k M\Phi(\mathbf{x})}
  \int_{0}^{2\pi}
  \frac{\E^{\J (k R + n\psi_{1})}}{4\pi R}
  \E^{\J k M\Phi(\psi_{1})}
  \,\D\psi_{1},
\end{align}
where for compactness we write $\Phi(\mathbf{x}_{0})=\Phi(\psi_{1})$
with $\mathbf{x}_{0}=(a_{0}\cos\psi_{1}, a_{0}\sin\psi_{1}, 0)$ on the
source ring.
Our task now is to rewrite the exponential containing
the flow potential in a form suitable for analytical treatment.

We handle the non-uniform flow effect as a set of azimuthal modes
based on a Fourier series for the flow potential at the source
positions,

  \begin{align}
    \label{equ:analysis:fourier}
    \Phi(\psi_{1}) 
    &=
    \Phi_{0} + 
    \sum_{q=1}^{\infty} 
    \Phi_{q}\E^{\J q\psi_{1}}
    +
    \Phi_{-q}\E^{-\J q\psi_{1}},
  \end{align}
and

  \begin{align}
    \E^{\J kM\Phi(\psi_{1})}
    &=
    \E^{\J k M\Phi_{0}}
    \E^{\J k M[\Phi(\psi_{1})-\Phi_{0}]},
  \end{align}
which to first order in $M$ can be approximated

  \begin{align}
    \E^{\J kM\Phi(\psi_{1})}
    &\approx
    \E^{\J k M\Phi_{0}}
    \left(
      1 + \J k M
      \sum_{q=1}^{\infty} 
      \Phi_{q}\E^{\J q\psi_{1}}
      +
      \Phi_{-q}\E^{-\J q\psi_{1}}
    \right)\nonumber\\
    &=
    \E^{\J k M\Phi_{0}}
    \sum_{q=-\infty}^{\infty}
    c_{q}\E^{\J q\psi_{1}},\\
    c_{0} &= 1,\\
    c_{q} &= -\J k M\Phi_{q},\quad q \neq 0,
  \end{align}
noting that $c_{-q}\neq c_{q}^{*}$. 

Given a source of azimuthal order $n$, the acoustic potential at some
field point $\mathbf{x}$ is then
\begin{align}
  I_{n}(k, M, a_{0}; r, x)
  &=
  \E^{-\J k M\Phi(\mathbf{x})}
  \E^{\J n\psi}
  \int_{0}^{2\pi}
  \frac{\E^{\J (k R + n\psi_{1})}}{4\pi R}
  \E^{\J k M\Phi(\psi_{1})}
  \,\D\psi_{1}\nonumber\\
  \label{equ:analysis:expansion}
  &= 
  \E^{\J k M[\Phi_{0}-\Phi(\mathbf{x})]}
  \sum_{q=-\infty}^{\infty}
  c_{q}\E^{\J (n+q) \psi}
  I_{n}(k, 0, a_{0}; r, x),
\end{align}
The integrals on the right hand side can be determined using the exact
series expression of Equation~\ref{equ:analysis:series}.

\subsection{Summary of method}
\label{sec:analysis:summary}

The calculation method for the problem of a ring source of order $n$
radiating into a non-uniform flow of Mach number $M$ can be
summarized:
\begin{enumerate}
\item compute the flow potential $\Phi(\mathbf{x})$ for unit velocity;
\item compute the coefficients of the Fourier series of
  $\Phi(\mathbf{x})$ on the source ring;
\item use the series of Equation~\ref{equ:analysis:series} and the
  summation of Equation~\ref{equ:analysis:expansion} to determine the
  acoustic field.
\end{enumerate}
In the general case, the Fourier coefficients of $\Phi(\mathbf{x})$
may have to be determined numerically, but in the next section we
present a model problem for which the required terms can be evaluated
analytically. 

\section{Model problem}
\label{sec:model}

The method as presented so far is quite general as long as the Fourier
expansion of the flow potential can be determined. In practice, this
will have to be done numerically for many problems, but in the case of
a source interacting with a cylinder, a model problem for a rotor near
a wing, the Fourier coefficients can be found analytically. The
problem contains the essential features of the modification of an
acoustic field by a realistic potential flow and it will be seen that
the flow gives rise to noticeable effects on the overall field, even
though it is a very simplified version of the problem which arises in
applications. 

\subsection{Geometry and notation}
\label{sec:model:geometry}

\begin{figure}
  \centering
  \includegraphics{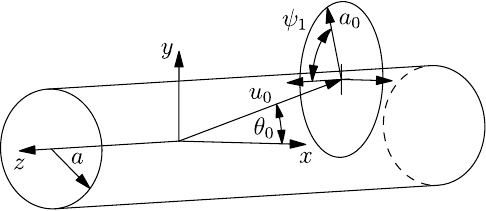}
  \caption{Geometry and notation for ring source of radius $a_{0}$
    radiating into potential flow around infinite cylinder of radius
    $a$}
  \label{fig:model:geometry}
\end{figure}

The model problem is shown in Figure~\ref{fig:model:geometry}. A wing
is represented by an infinite cylinder of radius $a$ lying along the
$z$ axis. The source of radius $a_{0}$ is placed with its axis
parallel to the $x$ axis and its center at $(x_{0},y_{0})$. In
calculating the flow potential, cylindrical coordinates $(u,\theta,z)$
are used. In this coordinate system, the source center lies at
$(u_{0},\theta_{0})$ and points on the ring are given by $(x_{0},
y_{0}+a_{0}\sin\psi_{1}, a_{0}\cos\psi_{1})$. 

\begin{figure}
  \centering
  \includegraphics{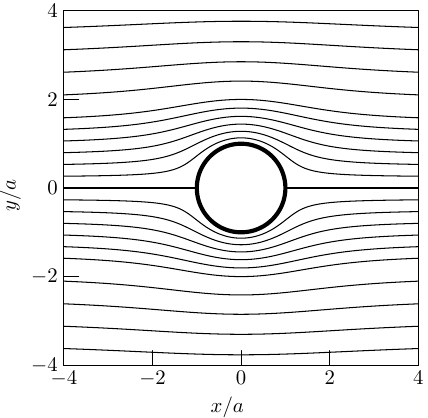}
  \caption{Streamlines for flow in positive $x$ direction around a
    cylinder of radius $a$.}
  \label{fig:model:streamlines}
\end{figure}

The solution for potential flow over a cylinder of radius $a$ can be
found in elementary fluid mechanics textbooks and in the notation of
this paper is
\begin{align}
  \label{sec:model:flow}
  \Phi(\mathbf{x}) 
  &= 
  x
  \left(
    1 + a^{2}/u^{2}
  \right),
\end{align}
for flow of unit velocity in the positive $x$
direction. Figure~\ref{fig:model:streamlines} shows the streamlines
for flow over a cylinder. The non-uniformity of the flow is clear near
the cylinder, and it is also clear that the flow is nearly uniform
more than about one cylinder radius from the cylinder surface.

In order to evaluate the effect of this flow on radiation from the
ring source, we expand $\Phi$ in a Fourier series in $\psi_{1}$, the
angular coordinate on the ring source of
Figure~\ref{fig:model:geometry}:
\begin{align}
  \Phi(\psi_{1})
  &=
  x_{0}
  \left(
    1
    +
    \frac{a^{2}}{u_{0}^{2} + 2a_{0}y_{0}\sin\psi_{1} + a_{0}^{2}\sin^{2}\psi_{1}}
  \right)\nonumber\\
  &= \sum_{q=-\infty}^{\infty}\Phi_{q}\E^{\J q \psi_{1}},\\
  \Phi_{q}
  &=
  \frac{1}{2\pi}
  \int_{0}^{2\pi}
  \Phi(\psi_{1})\E^{-\J q\psi_{1}}
  \,\D\psi_{1}.
\end{align}
Coefficients of the Fourier series can be evaluated using tabulated
integrals~\cite{gradshteyn-ryzhik80},
\begin{align}
  \Phi_{0} &= 
  x_{0} + \frac{a^{2}}{Ba_{0}}\cos(\gamma-\theta_{0}),\\
  \Phi_{q} &= 
  (-\J)^q\frac{a^{2}}{Ba_{0}}
  \left(
    B^{2} - 2BS\cos\gamma + S^{2}
  \right)^{q/2}
  \cos
  \left[
    (q-1)(\theta_{0}-\delta)
    + \gamma - \delta
    - q\pi/2
  \right],\, q>0,\\
  \Phi_{-q} &= \Phi_{q}^{*},
\end{align}
where
\begin{align}
  B &= 
  \left(
    S^{4} + 2S^{2}\cos2\theta_{0} + 1
  \right)^{1/4},\\
  S &= u_{0}/a_{0},\\
  \gamma &=
  \frac{1}{2}\tan^{-1}\frac{\sin2\theta_{0}}{S^{2}+2\cos2\theta_{0}},\\
  \delta &=
  \tan^{-1}\frac{B\sin\gamma}{B\cos\gamma - S}.
\end{align}
The acoustic field of the ring source in flow over the cylinder can
then be evaluated by substituting $\Phi_{q}$ into
Equation~\ref{equ:analysis:expansion}. 

\subsection{Results}
\label{sec:model:results}

\begin{figure}
  \centering
  \includegraphics{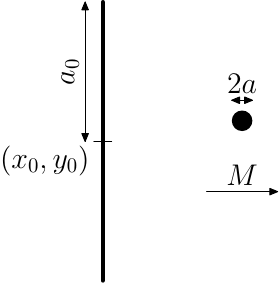}
  \caption{Geometry of rotor position relative to cylinder in mean
    flow for parametric study}
  \label{fig:model:results:geometry}
\end{figure}

\begin{figure}
  \centering    
  \includegraphics{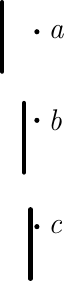}     
  \caption{Rotor positions for parametric study. Case \textit{a}:
    $(x_{0},y_{0})=(-a_{0},-a)$; \textit{b}:
    $(x_{0},y_{0})=(-5a/2,-a_{0}/2)$; \textit{c}:
    $(x_{0},y_{0})=(-5a/4,-a_{0}/2)$.}
  \label{fig:model:results:positions}
\end{figure}

To apply the results of the paper to a representative problem,
parameters have been estimated for the ATR class of turboprop
aircraft. The propeller diameter and speed are~3.93~\meter\
and~1200~rpm respectively, so that the rotation Mach number for the
source is~0.73. The cylinder diameter has been set equal to twice the
maximum thickness of the ATR wing, with $a=0.29~\meter$.  Flow Mach
number $M=0.15$, a typical value for takeoff of a turboprop
aircraft. Results have been calculated for a baseline case of $n=6$,
representing the lowest harmonic of the tonal noise from a six-bladed
propeller. Figure~\ref{fig:model:results:geometry} shows the notation
for the rotor position relative to the cylinder, while
Figure~\ref{fig:model:results:positions} shows the configurations
selected for the parametric study. The source is placed at negative
$x_{0}$ to correspond to a rotor ahead (upstream) of a wing leading
edge.

\begin{figure}
  \centering
  \includegraphics{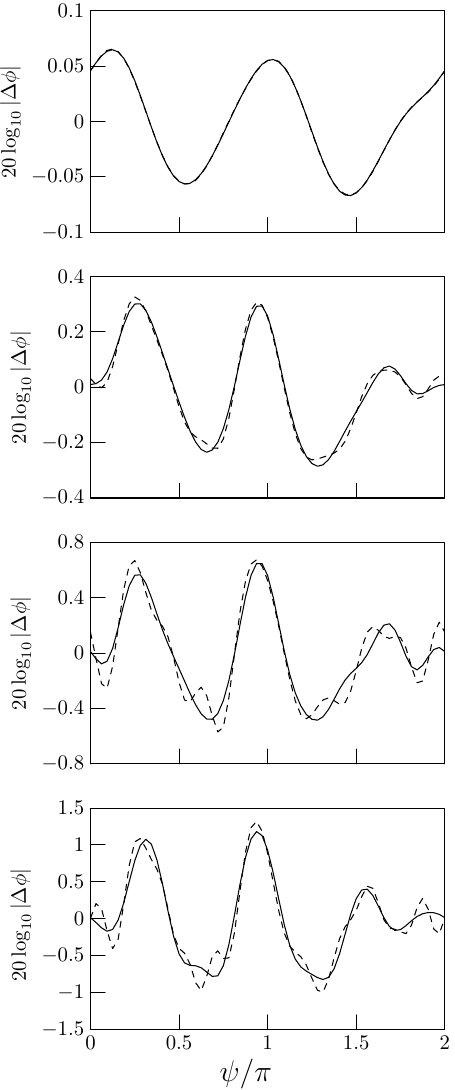}
  \caption{Magnitude of computed acoustic potential with $M=0.15$
    flow, scaled on no-flow result. From top to bottom: case
    \textit{a}, \textit{b}, \textit{c}, and \textit{c} with
    $n=8$. Solid line analytical result, dashed line numerical
    evaluation.}
  \label{fig:model:results}
\end{figure}

Sample calculations for the problem are presented in
Figure~\ref{fig:model:results}, in the form of the magnitude of
acoustic potential with flow, scaled on the no-flow case. As a check
on the method, the integral of Equation~\ref{equ:analysis:taylor:int}
was evaluated numerically and its magnitude is also plotted. Results
are shown for $n=6$ in each of the configurations of
Figure~\ref{fig:model:results:positions} and for $n=8$ in the third
configuration. In each case the field is calculated in the source
plane on a circle of radius $r=2a_{0}$. In considering the magnitude
of flow effects on the radiated noise, it is useful to note that
Hanson~\cite{hanson95} found that a flow of Mach number $M=0.2$
at~10\degree\ to the rotor axis altered the sound level by
about~1.5\deci\bel. In our case, the flow is nominally at~0\degree\
and all asymmetry in the system arises from the flow non-uniformity.

The first plot, for a rotor some distance upstream of the cylinder,
shows a very small change in the acoustic potential. This is as might
be expected from Figure~\ref{fig:model:streamlines} since the rotor
lies in a region of flow which is almost uniform and very little
asymmetry will be introduced by the changes in potential around the
source ring. Moving the source closer to the cylinder, however,
generates quite large changes in the acoustic field which, we
emphasize, are purely the result of changes in the radiation
properties of the source, which is held constant throughout the
calculations. In the second plot, with the source brought within two
radii of the cylinder surface, the field is modified by up
to~0.3\deci\bel. In the third plot, moving the source closer still, a
quarter radius from the cylinder surface, generates a larger change in
the acoustic field, of some~0.6\deci\bel. The largest change shown,
however, comes with the rotor in the same position, close to the
cylinder, but with $n$ increased from~6 to~8, approximating the
replacement of a six-bladed propeller with an eight-bladed. The change
in the acoustic field is now greater than~1\deci\bel, comparable to
the effect of a flow at an angle of~10\degree\ in Hanson's study of
incidence effects. This effect is caused by the introduction of new
modes with higher $M_{t}$ radiating efficiently into the field.

\section{Conclusions}
\label{sec:conclusions}

An analysis has been developed for the prediction of sound from
rotating sources in a low Mach number non-uniform potential flow,
giving an exact formulation for the radiated field as a sum of
azimuthal modes whose amplitude is determined by the local flow.  The
theory has been applied to the model problem of a rotor operating in
the vicinity of a cylinder, for which the required coefficients can be
found analytically. Sample calculations compare well to numerical
evaluation of the acoustic field and the results show changes in the
field which vary strongly with rotor position relative to the
cylinder. The effect of flow non-uniformity on radiation properties
has been found to be comparable in magnitude to the effect of
incidence in propeller acoustics. Future work will consider disk
sources, which have finite radial extent, modeling propeller blades of
finite span, and the effects of wing lift on the radiated noise, as
finite circulation on the cylinder introduces changes in the flow
potential. An important open question remains that of the relationship
between the model problem and realistic systems which require
computational aeroacoustics methods for their prediction.


\section*{Acknowledgments}

The author is grateful to the reviewers and the editor, in particular
for noting the relationship between Taylor's transformation and the
geometric acoustics approximation.


\appendix
\section{Taylor's transformation as geometrical acoustics}
\label{sec:geometrical}

Taylor's transformation, Equation~\ref{equ:analysis:taylor}, was
originally derived using a coordinate transformation in the wave
equation~\cite{taylor78}. During the review process of this paper, it
was noted that the form of Equation~\ref{equ:analysis:taylor} suggests
that it might also be derived from the geometrical acoustics
approximation. This appears to be a novel result and may be useful in
developing other methods for the wave equation with mean flow. The
geometrical acoustics approximation gives the time delay for propagation
between two points as a function of the speed of sound and the fluid
velocity along the propagation path. Equation~8-1.14 of
Pierce~\cite{pierce89} gives the propagation time $T_{AB}$ between
points $A$ and $B$ of a ray as
\begin{align}
  \label{equ:geometrical:time}
  T_{AB}
  &=
  \int_{\ell_{A}}^{\ell_{B}}
    \frac{\D\ell}{\mathbf{v}\cdot\mathbf{x}' + [c^{2} - v^{2} +
      (\mathbf{v}\cdot\mathbf{x}')^{2}]^{1/2}},
\end{align}
where $\ell$ is distance along the ray, $\mathbf{v}$ is local fluid
velocity, and $\mathbf{x}'=\D\mathbf{x}/\D\ell$. Given that
$\mathbf{v}=cM\nabla\Phi(\mathbf{x})$,

  \begin{align}
    T_{AB}
    &=
    \frac{1}{c}
    \int_{\ell_{A}}^{\ell_{B}}
    \frac{\D \ell}
    {M\nabla\Phi\cdot\mathbf{x}' + [1-M^{2}|\nabla\Phi|^{2} +
      M^{2}(\nabla\Phi\cdot\mathbf{x}')^{2}]^{1/2}},
  \end{align}
which is, to first order in $M$,

  \begin{align}
    T_{AB}
    &=
    \frac{1}{c}
    \int_{\ell_{A}}^{\ell_{B}}
    1 - M\nabla\Phi\cdot\mathbf{x}'
    \,\D \ell\nonumber\\
    &=
    \frac{1}{c}
    \int_{\ell_{A}}^{\ell_{B}}
    1 - M\frac{\D\Phi}{\D\ell}
    \,\D \ell\nonumber\\
    &= \frac{1}{c}
    \left[\ell_{B} - \ell_{A} - M(\Phi_{B} - \Phi_{A})\right].
  \end{align}
Identifying point $A$ with the source position $\mathbf{x}_{0}$, point
$B$ with the field point $\mathbf{x}$, and $\ell_{B}-\ell_{A}$ with
the distance $R$, this gives the phase term of
Equation~\ref{equ:analysis:taylor}.




\end{document}